\documentclass[a4paper,12pt]{article}
\usepackage[utf8]{inputenc}
\usepackage{cancel}
\usepackage{tikz}
\usepackage{ulem}
\usepackage{amsfonts}
\usepackage{amssymb}
\usepackage{graphicx}
\usepackage{amsmath}
\usepackage{enumerate}
\usepackage{mathtools}
\usepackage{subfig}
\usepackage{color}
\usepackage{tikz}
\usepackage{float}
\usepackage{here}
\usepackage{cite}
\usepackage{mathrsfs}
\usepackage{float,epsfig}
\usepackage{dcolumn}
\usepackage{graphicx}
\usepackage{bm}
\usepackage{amsmath,amssymb,amsthm}
\usepackage[colorlinks=true,linkcolor=blue,citecolor=red]{hyperref}
\usepackage{multirow}
\usepackage[toc,page]{appendix}
\usetikzlibrary{arrows.meta}
\usetikzlibrary{bending}
\usetikzlibrary{calc}
\newcommand{\be}{\begin{equation}}
\newcommand{\ee}{\end{equation}}
\newcommand{\bea}{\setlength\arraycolsep{2pt} \begin{eqnarray}}
\newcommand{\eea}{\end{eqnarray}}

\def\0{{\sst{(0)}}}
\def\1{{\sst{(1)}}}
\def\2{{\sst{(2)}}}
\def\3{{\sst{(3)}}}
\def\4{{\sst{(4)}}}
\def\5{{\sst{(5)}}}
\def\6{{\sst{(6)}}}
\def\7{{\sst{(7)}}}
\def\8{{\sst{(8)}}}
\def\sst#1{{\scriptscriptstyle #1}}

\makeatletter \@addtoreset{equation}{section}

\definecolor{lime}{HTML}{A6CE39}

\begin{document}

\title{{\normalsize \textbf{\Large Hypergeometric Potential Inflation and Swampland Program in   Rescaled Gravity with Stringy  Corrections   }}}
\author{ {\small  Saad Eddine Baddis\footnote{saadeddine.baddis@um5r.ac.ma}    \;  and  Adil  Belhaj\footnote{a-belhaj@um5r.ac.ma}    \thanks{%
\bf Authors in alphabetical order.} \hspace*{-8pt}} \\
{\small D\'{e}partement de Physique, \'Equipe des Sciences de la
mati\`ere et du rayonnement, ESMaR} \\ {\small Facult\'e des Sciences, Universit\'e Mohammed V de Rabat, Rabat,
Morocco}}
\maketitle

\begin{abstract}
Motivated  by string theory activities,  we  investigate  inflationary models and   the  swampland criteria  in  the context of a  stringy rescaled gravity. 
 Inspired by differential equations associated with special functions, we develop an algorithm to  derive new    scalar potential functions with hypergeometric  behaviors from    string theory  correction terms.  Among  others,  we  obtain a family of  models  indexed by a couple  $(m,n)$, where $m$ and $n$ are natural numbers  constrained    by  hypergeometric  behaviors and  certain physical requirements.  Using the falsification scenario,  we  confront  the derived   models with the Planck  observational data for  such a   stringy rescaled gravity.  Then, we   approach  the  associated swampland conjectures.   For certain models of phenomenological interest, we find that   the swampland criteria are satisfied for small values of  the slow-roll   parameters in such a modified gravity.

\end{abstract}

\newpage

\section{Introduction}
Recently,  the  swampland criteria program   has received  a remarkable interest in connection with various  theories  including the high energy physics one \cite{R1,R11,R2,R3,R4,R5}. This program   seems to have a great effect on how to  approach effective field theories (EFT's)  and thier  consistency with quantum gravity.  In fact, it allows one to distinguish between EFT's that couple consistently with quantum gravity  and  the ones  that are inconsistent  with  such a  coupling. It has been motivated by non-trivial  gravity theories  including black holes and string theory \cite{ST1,ST2,ST3}. Moreover, it has been suggested that 
the swampland criteria has been developed   also in connection with  the dark dimension  \cite{D1,D2,D3}.   More recently, a  particular emphasis  has been put on its application to    inflation phenomenon. The   inflation scenario  has been    exploited  to overcome     many issues concerning the standard cosmology, including   the horizon and  the flatness  problems  \cite{I1,I11,I22,I33,I2,I3,I4,I5,B1,B2}. Moreover,   it  has been extensively   investigated  in relation  with other topics such as black holes,  dark energy and dark matter \cite{BH1,DAM1,DAE1}. In this context,  it has been observed the importance of   the scalar fields providing  certain physical features  associated with
homogeneity and isotropy properties. These objects can be embedded in  several physical theories, including string theory. In this  theory and related topics,  the scalar fields are generally derived from the  compactification mechanism  on non-trivial spaces, such as   Calabi-Yau  and G2 manifolds\cite{IST1,IFM1,IFM2}.  However, the simplest model involves a single homogeneous scalar field,  and its potential,  which  could interact with gravity in various ways.   Several  forms of the scalar potential  have been proposed  in order to build inflationary models from different theories of gravity, including the  modified ones \cite{MD1,MD2,MD11,MD22,MD3,MD4,B003,B3,B4,B5}.  However,  certain of such theories have been highlighted and omitted  by observational  findings from Planck data, where the relevant cosmological   quantities do not lie within the observational   ranges \cite{O1,O2,O3}.  In order to find models in a  good agreement with such  data,   a close inspection shows  that numerous  approaches and roads  have been  suggested.  In connection with the CDM model \cite{CDM}, the analysis of  inflationary models has  been  elaborated by considering
perturbation parameters. In addition, gravitational models inspired by string  theory  and M-theory compactifications have been also 
 studied to describe inflation.  In this way,  the scalar field  can be  linked to the  geometric deformations of the internal spaces  controlled  either  by  the size   or  the shape parameters  \cite{CY1,CY2}. In these activities, several  models  involving  a scalar field have been studied to  bridge  the  theoretical predictions  with
the observational data provided by the Cosmic Microwave Background (CMB) and the Planck 
experimental results \cite{O1,O2,O3}.

 A close examination  reveals that the  inflationary models  derived from  modified  general relativity (MGR) have been also  explored   showing interesting results \cite{OS1,OS2,OS3,OS4,OS5,OS6}. Precisely,  the most widely dealt with  models are $f(R)$ modified  gravities,  where  $R$ is the Ricci scalar.  One simple  possible way to go beyond  the  general relativity (GR)
 is to  consider   $f(R)=\alpha R $ generating   a rescaled Einstein-Hilbert gravity  \cite{RG1,RG2,RG3,RG4,RG5,3}. Using the  slow-roll  approximations,  relevant cosmological  quantities such as the spectral index $n_S$  and the the tensor/scalar ratio $r$  have been   computed  and examined  via  different  scalar potentials  for a given number of  the e-foldings supported by empirical results. Certain  models have provided  numerical values in  a perfect agreement with 
the Planck collaboration results.  Motivated by  string theory, other scenarios  have been explored producing  corroborated  models. Specifically, these include the Gauss-Bonnet gravity theories  where  the cosmological quantities have been determined and discussed in many places.    Recently, the  stringy corrections have been implemented  to such a rescaled gravity   via 
a  scalar  function denoted by  $\xi=\xi(\phi)$  describing  the coupling between the scalar field $\phi$ and the
Gauss-Bonnet invariant term.    In such a rescaled Einstein-Hilbert gravity theory, the  stringy corrections   have been dealt with in connection   with  the swampland criteria program.  

The aim of this paper is to study   the swampland program of   certain  inflationary models in the context of a   stringy rescaled gravity. 
 Inspired by differential equations associated with special functions, we develop an algorithm to derive new   scalar potential functions with hypergeometric  behaviors from    string theory  correction terms.   Precisely, we   expose a family of    models  indexed by a couple  $(m,n)$, where $m$ and $n$ are natural numbers  constrained    by  hypergeometric  behaviors and physical requirements.  Using the falsification scenario,  we  confront  the derived   models with the Planck  observational data for  such a   stringy rescaled gravity.  Then, we   discuss   the  associated swampland conjectures.   For certain models of phenomenological interest, we find that   the swampland criteria are satisfied for small values of  the slow-roll   parameters in such a  modified gravity.

The organization of this paper is as follows. In  section 2, we  elaborate a concise presentation  on the proposed   rescaled gravity  theory  and   the   swampland criteria  scenarios.  In section 3,   we provide an algorithm to construct   inflationary models from  inspired string theory corrections. In section 4,  we provide an artwork for   the present analysis by checking  the validity of the
swampland criteria for certain models of phenomenological interest. The last section is devoted to conclusions and open questions.

\section{Stringy rescaled gravity with  swampland program}
\label{sec:2}
In this section, we reconsider the study of  a family of models in a  stringy rescaled gravity.  The latter is described by 
a  minimal Einstein-Hilbert  contribution  with some of sets of corrections motivated  by string theory.  This involves couplings  
of a  scalar field  to the Einstein tensor   and the   4-dimensional  Gauss-Bonnet (GB)  term via differentiable functions. The introduction of these coupling functions has been  presented   in many places including  \cite{3,2}, where they have been considered as a measure to the contribution of high order curvature terms, also known as corrections to the Einstein-Hilbert action. Taking  $M_p^{-2}=8\pi G=1$,   the corresponding   action, in the Einstein frame,  can be  expressed as follows
\begin{equation}
\label{e0021}
S=\int d^4x\sqrt{-g}\left[\frac{\alpha }{2\kappa^2}R-\frac{1}{2}\partial_{\mu}\phi\partial^{\mu}\phi-V(\phi)
+\zeta(\phi)G_{\mu\nu}\partial^{\mu}\phi\partial^{\nu}\phi-\xi(\phi){\cal G}\right]
\end{equation}
{where  $R$ is the Ricci scalar}.  $\phi$ is a real scalar field with a
potential  function $V(\phi)$.   $G_{\mu\nu}$ is the Einstein  tensor,  and  ${\cal G}$  represents  the GB 4-dimensional invariant  term given by ${\cal G}=R^2-4R_{\mu\nu}R^{\mu\nu}+R_{\mu\nu\lambda\rho}R^{\mu\nu\lambda\rho}$. It  is worth noting $\zeta(\phi) $ and $\xi(\phi)$ are two real   differentiable functions of $\phi$   describing the stringy corrections to the rescaled gravity.  The parameter $\alpha$ basically describes the dominant term in the $f(R)$ gravity. In the case of  the Gogoi-Goswami gravity, this  $f(R)                                                  $ function is  given by \begin{equation}
 f(R)=R-\frac{c_1}{\pi}R_c \cot^{-1}(\frac{R_c^2}{R^2})-c_2R_c(1-e^{-\frac{R}{R_c}})
 \end{equation}
 where  $R_c$ is  a characteristic curvature.  $c_1$ and  $c_2$ are dimensionless parameters.\\ 
This type of gravity has been   extensively investigated in many places including   \cite{GG1,GG2}. 
  At the early inflationary era,  we  could consider large values of   $R$   leading to the asymptotic behavior
 \begin{equation}
 f(R)\approx R(1-c_2)=\alpha R.
 \end{equation}
 According  to \cite{GG2}, the  $c_2$ parameter is constrained by $0<c_2<1$, needed to avoid certain  tachyonic instabilities. Moreover, the small values of $c_2$ are the most acceptable ones in the solar system test within  the Jordan frame \cite{J1}.  This small value behavior on $c_2$ is translated to a condition on $\alpha$ being $0<\alpha<1$. This parameter could  be absorbed in $\kappa$, but for representation sake, it is convenient to work in the range ]0,1[. To  be consistent with the naturalness argument, however, it is reasonable to treat $\alpha$ as a coupling parameter.

 The above action provides a  family of models  
depending on  three  scalar   functions  $V\phi)$, $\zeta(\phi)$ and $\xi(\phi)$.  A priori these functions  should be arbitrary.  To elaborate  corroborated models 
 matching with   the observational  data  \cite{4,5,6,7,8,9,10,11},  however, certain requirements should be imposed on such scalar functions.  
 Using $\phi$ and  $g_{\mu\nu}$ variations, we can obtain  the equations of motion by means of   the Freedman-Robertson-Walker metric
  \begin{eqnarray}
  ds^2=-dt^2+a(t)(dx^2+dy^2+dz^2)
   \end{eqnarray}
 where $a(t)$ is the scale factor  describing  the Universe evolution.   Indeed, the equations  of motion are found to be 
 \begin{eqnarray}
\label{e0025}
\frac{3\alpha H^2}{\kappa^2}&=&\frac{1}{2}\dot{\phi}^2+V(\phi)+9H^2\dot{\phi}^2\zeta(\phi)+24H^3\dot{\xi}(\phi) \\
\label{e0026}
-\frac{2\alpha\dot{H}}{\kappa^2}&=&\dot{\phi}^2+
\big[(6H^2-2H)\zeta(\phi)-2H\dot{\zeta}(\phi)\big]\dot{\phi}^2-16\dot{H}H\dot{\xi}(\phi)\notag \\
&-& 8H^2(\ddot{\xi}(\phi)-H\dot{\xi}(\phi))-4H\zeta(\phi)\ddot{\phi}\dot{\phi}\\
\label{e0027}
\ddot \phi  + 3H\dot \phi  + V'(\phi)&=&
- 24{H^2}\left( {{H^2} + \dot H} \right)\xi^\prime(\phi)
- 6H\zeta(\phi)\dot \phi (3{H^2} + 2\dot H)\nonumber\\ &&
- 3{H^2}\left( 2\zeta(\phi)\ddot \phi  + \zeta^\prime(\phi)\dot \phi ^2 \right),
\end{eqnarray}
where the  prime  is the derivative with respect to the scalar  field $\phi$ and  the dot  is the derivative with respect to  the time.  $H$ denotes the Hubble parameter  defined by $H=\frac{\dot{a}}{a}.$ These equations of motion can recover known certain models.  Ignoring  the kinetic Einstein coupling,    for instance, such  equations reduce to
\begin{eqnarray}
\frac{3\alpha H^2}{\kappa^2} &=& \frac{1}{2}\dot{\phi}^2+V(\phi)+24\dot{\xi}(\phi)H^3,\\
-\frac{2\alpha \dot{H}}{\kappa^2} &=& \dot{\phi}^2-16\dot{\xi}(\phi)H\dot{H}-8H^2(\ddot{\xi}(\phi)-H\dot{\xi}(\phi)),\\
\ddot{\phi}+3H\dot{\phi}+V^\prime(\phi)+\xi^\prime(\phi)\mathcal{G} &=& 0.
\end{eqnarray}
A close examination shows that the scalar  field and its potential should satisfy   appropriate  requirements.  The most investigated ones concern   the swampland criteria being  elaborated in \cite{R11} to be in tension with the inflationary theory.  Specifically,  it has been  revealed  that the de Sitter conjectures are incompatible with the  ranges of the slow-roll indices for positive potentials.  This  is due to the fact that the slow-roll indices  with  large values lead  to the problematic of initially fine-tuned conditions.
  However,   it could be possible to  show  that this is not the case for certain  models.  In this way,  we could still extract models that meet all the requirements put by the inflationary constraints, to fit the theory in the bounds of the Planck data, and the swampland criteria. This  could be  approached  by checking these constraints for different sets of the free parameters of the considered  models.  The  exploit of  the swampland program  allows  one to add more constraints on the proposed models, narrowing the possible values of their free parameters.  One can also discern if the models has a stringy underlining or not.    Roughly, the  additional constraints manifest from 
\begin{itemize}
\item the swampland distance conjecture being
\begin{equation}
\mid \kappa\Delta\phi\mid<\mathcal{O}(1),
\end{equation}
\item the de sitter conjectures which  are 
\begin{eqnarray}
\frac{\mid V^\prime(\phi)\mid}{\kappa V(\phi)}>\mathcal{O}(1), \qquad 
-\frac{V^{\prime\prime}(\phi )}{\kappa^2V(\phi)}>\mathcal{O}(1).
\end{eqnarray}
\end{itemize}

\section{Inflationary algorithm for the  stringy rescaled gravity}
To  elaborate  models which could be corroborated via the falsification  mechanism, we need to provide possible  inflationary    predictions. Indeed,  
the previous  equations of motion  can be approached   via the slow-roll  analysis by imposing physical  requirements. During  the  inflation phase,  one can expose the slow-roll parameters  as  follows  \begin{eqnarray}
\label{Eq3.1}
\epsilon_1 &=& -\frac{\dot{H}}{H^2},\\
\label{Eq3.2}
\epsilon_2 &=& \frac{\ddot{\phi}}{H\dot{\phi}},\\
\epsilon_3 &=& \frac{\dot{E}}{2HE},\\
\epsilon_4 &=& \frac{\dot{Q}_{GB}}{2HQ_{GB}},
\end{eqnarray}
where one has used  ${Q}_{GB}=\frac{\alpha}{\kappa^2}-8\dot{\xi}(\phi)H$ and $E=\frac{\alpha}{\kappa^2}(1+\frac{3Q^2_a}{2Q_{GB}\dot{\phi}^2})$ with $Q_a=-8\dot{\xi}^2(\phi)$.
Following  the slow-roll inflation aspect,  such  parameters  are conditioned by 
\begin{equation}
	\epsilon_1, \epsilon_2, \epsilon_3,\epsilon_4 <<1.
\end{equation}
The ultimate goal here is to construct an  inflationary cosmological models in the  swampland  program  of  the rescaled gravity theories, whose  the viability depends on the  observational constraints on the relevant indices being, the scalar spectral index of primordial perturbations $n_S$, the tensor spectral index   $n_\tau$  and the tensor to  the scalar ratio $r$. In addition,  one needs to determine the numerical range of the parameter $\alpha$, and the  remaining parameters of the models, being compatible with the swampland criteria. Roughly speaking, we introduce the slow-roll conditions
\begin{equation}
\dot{H}\ll H^2,\hspace{10 pt}\frac{1}{2}\dot{\phi}^2\ll V(\phi),\hspace{10 pt}\ddot{\phi}\ll 3H\dot{\phi}.
\end{equation}
The Gauss-Bonnet scalar coupling function $\xi(\phi)$ contribution affects the velocity of the propagation of the tensor perturbations, where the primordial gravitational waves are no longer constrained to propagate with the light velocity. The velocity in this case is given by the expression
\begin{equation}
c^2_\tau=1-\frac{Q_f}{Q_{GB}},
\end{equation}
where  $Q_f=8(\ddot{\xi}(\phi)-H\dot{\xi}(\phi))$  is an auxiliary  quantity, with $\dot{\xi}(\phi)$ is the derivative of $\xi(\phi)$ with respect to time.  To generate a  compatibility  with the GW170817 event \cite{4,5,6,7,8,9,10,11},  and more recently the observations of EPTA, Parkes Observatory and CPTA \cite{12,13,14,15,16}, the Gauss-Bonnet scalar coupling function should be a solution of $\ddot{\xi}(\phi)=H\dot{\xi}(\phi)$   considered as    a differential equation form  being  shown in \cite{17,18,19}. Thus, the velocity is still the unity and the theory still preserves  the causality. Using the canonical expansion $\dot{\xi}(\phi)=\xi^\prime(\phi)\dot{\phi}$,  one gets
\begin{equation}
\ddot{\phi}\xi^\prime(\phi)+\xi^{\prime\prime}(\phi)\dot{\phi}^2=H\dot{\phi}\xi^\prime(\phi)
\end{equation}
where  $\xi^\prime(\phi)$ is the derivative of $\xi(\phi)$ with respect to the scalar field.
Applying the slow-roll conditions, we obtain the time derivative of the homogeneous  scalar field
\begin{equation}
\dot{\phi}\simeq\frac{H\xi^\prime(\phi)}{\xi^{\prime\prime}(\phi)}.
\end{equation}
Combining the constraints on the speed of the gravitational waves and the slow-roll conditions, the equations of motion reduce to
\begin{eqnarray}
\frac{3\alpha H^2}{\kappa^2} &\simeq & V(\phi)+24\dot{\xi}(\phi)H^3,\\
-\frac{2\alpha \dot{H}}{\kappa^2} &\simeq &\dot{\phi}^2-16\dot{\xi}(\phi)H\dot{H},\\
3H\dot{\phi}+V^\prime(\phi)+24\xi^\prime(\phi) H^4 &\simeq & 0.
\end{eqnarray}
To extract information on the inflationary phenomenology, further approximations should be imposed.  Specifically, we neglect the string  theory corrections due to their  small and practically vanishing numerical contributions. However,  the stringy information still survive in the time derivatives of the scalar field. The  equation system takes the following form
\begin{eqnarray}
\frac{3\alpha H^2}{\kappa^2} &\simeq & V(\phi),\\
-\frac{2\alpha \dot{H}}{\kappa^2} &\simeq & \left(\frac{H\xi^\prime(\phi)}{\xi^{\prime\prime}(\phi)}\right)^2,\\
V^\prime(\phi) +\frac{\xi^\prime(\phi)}{\xi^{\prime\prime}(\phi)}\frac{\kappa^2V(\phi)}{\alpha}+\frac{8}{3\alpha^2}\xi^\prime(\phi)\kappa^4V^2(\phi)  &\simeq & 0.
\end{eqnarray}
In this system, the considered indices give approximated values to the scalar spectral index of the  primordial perturbations, the tensor spectral index and the tensor to scalar ratio. Indeed, they are given by 
\begin{eqnarray}
n_S &\simeq  & 1-2(\epsilon_1+\epsilon_2+\epsilon_3),\\
n_\tau &\simeq & -2(\epsilon_1+\epsilon_4),\\
r &\simeq & \vert 16\left(\alpha\epsilon_1-\frac{\kappa^2Q_e}{4H}\right)\frac{c^3_\mathcal{A}}{\kappa^2Q_{GB}}\vert.
\end{eqnarray}
The field propagation velocity is given by $c_\mathcal{A}=1+\frac{Q_aQ_e}{2Q_{GB}\dot{\varphi}^2+3Q^2_a}$, with an additional auxiliary parameter $Q_e=-32\dot{\xi}(\phi)\dot{H}$. These observational indices are evaluated during the first horizon crossing. According to the Planck data \cite{16}, the cosmological constraints are $n_S=0.9649\pm 0.0042$ and $r<0.064$. However, the constraint on the  tensor index $n_\tau$ is not yet determined. Using the  stringy corrections,  however,   it has been revealed that  constraints on such an index could be  imposed \cite{17,2}.\\
The initial value of the inflation is extracted from the e-folding number, using the formulated expression of $\dot{\varphi}$.  Indeed,  it  reads as
\begin{equation}
N=\int^{t_f}_{t_i}Hdt=\int^{\phi_f}_{\varphi_i}\frac{H}{\dot{\phi}}d\phi=\int^{\varphi_f}_{\varphi_i}\frac{\xi^{\prime\prime}(\phi)}{\xi^\prime(\phi)}d\phi.
\end{equation}
Usually,  this has been  evaluated by assumption to be in the interval  range $[50,60]$.  At this level, we would like to provide some comments. First, the ratio  $\frac{\xi'}{\xi''} $  appears almost in all equations.  This ratio should  deserve  its  importance. The second comment concerns the scalar potential   being  constrained by several programs  including  the swampland criteria. A close examination reveals   that  the above  ratio could be exploited to provide an algorithm allowing one  to elaborate a generic investigation for  the scalar potential forms.  Instead of considering particular forms,  we  can anticipate the existence of a differential equation provided by the string theory  correction functions.  Inspired by a  related investigation work \cite{3}, the stringy Gauss-Bonnet correction could verify the  following differential equation 
\begin{equation}
L_{\phi}\xi(\phi)=0
\end{equation}
where $L_{\phi}$ is a differential operator depending on  the scalar  field $\phi$    expressed as follows
\begin{equation}
L_{\phi}=A_{2}(\phi)\frac{d^{2}}{d\phi^{2}}+ A_{1}(\phi)\frac{d}{d\phi}+A_0(\phi).
\end{equation}
Equivalently, we can write 
 \begin{equation}
	A_2(\phi) \xi''(\phi)+A_1(\phi)\xi'(\phi)+ A_0(\phi)\xi(\phi)=0
\end{equation}
 with  two constraints $\xi(0)=0$ and $\xi(\infty)=-1$.   $A_i(\phi)$  are real functions which could be  associated with the equations of motion.  It has been remarked that      $A_0(\phi)$ could   be  linked  to   the    stringy correction function  $\zeta(\phi)$ corresponding to   the kinetic Einstein coupling ignored in the present investigation.   In mathematical language, this  type of differential  equations  has  been dealt with in connection with special  functions, playing a relevant role in quantum  physics.  A  generic  study may need more reflections. However, we consider only the following form
\ \begin{equation}
	A_2(\phi) \xi''(\phi)+A_1(\phi)\xi'(\phi)=0
\end{equation}
by neglecting  the     stringy correction $\zeta(\phi)$.    In this way,  the above equation can be solved by 
\begin{equation}
\frac{\xi'(\phi)}{\xi''(\phi)} =-  \frac{A_2(\phi)}{A_1(\phi)}.
\end{equation}
Handling this differential solution,   we obtain  a  general form for  the  stringy scalar coupling function being given by
\begin{equation}
\xi(\phi)=\lambda\int^\phi_0 e^{-g(x)}dx
\end{equation}
where $\lambda=-\xi^\prime(0)$ is the coupling constant. To be conformed with the boundary conditions,  one must have
\begin{equation}
\xi^\prime(0)=\frac{1}{\int^\infty_0 e^{-g(x)}dx}
\end{equation}
where the boundary conditions contribute  positively  to the naturalness of  EFTs. The function $g(x)$ is expressed as follows
\begin{equation}
g(x)= \int^x_0  \frac{A_1(y)}{A_2(y)}dy.
\end{equation}
In this algorithm,  the  scalar  potential is found to  be like 
\begin{equation}
	V(\phi)=  \frac{V_1(\phi)}{V_2(\phi)}
\end{equation}
where   $V_1(\phi)$ and $V_2(\phi)$ are  two real scalar  functions  given by 
\begin{eqnarray}
V_1(\phi) &=&   e^{f(\phi)}\\
V_2(\phi) &= & c+  \lambda \frac{8\kappa^2}{3\alpha} \int^\infty_{\phi} e^{h(x)}dx
\end{eqnarray}
{ where $c$  is an  integration constant  which can be treated as the inverse squared  of the cosmological constant $[c]=[M_p]^4$}. $f(x)$ and $h(x)$ are real functions being  expressed respectively as follows
\begin{eqnarray}
    f(x) &=& \int^\infty_{x}  \frac{A_2(y)}{A_1(y)}dy,\\
	h(x) &=& g(x)+f(x).
\end{eqnarray}
 Having elaborated an algorithm to  provide scalar potentials from  string theory  corrections, we    investigate  special models in the next section.  To insure the convergence of  the present solutions, we have to   impose  extra  conditions. Concretely,   it has been checked that   the  arbitrary differential function  $ \frac{A_2(\phi)}{A_1(\phi)}$   must adhere to the following constraint 
 \begin{equation}
   \lim\limits_{\phi\to + \infty}    \frac{A_2(\phi)}{A_1(\phi)}  =0.
\end{equation}
In what follows, such  differential functions will be relevant in the present investigation to build  models of phenomenological interest  with acceptable predicted numerical values. 
Instead of being general, we will consider certain function forms. The general studies could be developed in future works.
\section{$(m,n)$-models}
In this section, we  would like to provide certain family models involving new scalar potentials with hypergeometric behaviors. More precisely, we present a family  of models of  phenomenological interest using special functions including the   hypergeometric ones.  
Roughly, we propose    models relaying on the following  functions 
 \begin{eqnarray}
 A_2(\phi) &=& \beta  \\
A_1(\phi) &= & m(\kappa\phi)^{m-1}((\kappa\phi)^n+1) 
\end{eqnarray}
where $\beta$ is  a coupling parameter.  $m$ and $n$ are now arbitrary numbers.  We refer  to  as  $(m,n)$-models. The reason behind the choice of the set of  the functions $A_1$ and $A_2$ is to provide accessible forms of the scalar coupling function by avoiding large number of free parameters in the resulting theory. This could be a fruitful and an efficient path to probe the physical quantities of interest.
Roughly speaking, the proposed form of the differentiable functions could serve only to explore the swampland aspects in the theory in a more systematic manner. The main idea of this section is to present a methodical approach based on a differential algebraic constraint.\\

\begin{center}
\begin{figure}
\centering
\includegraphics[scale=0.58]{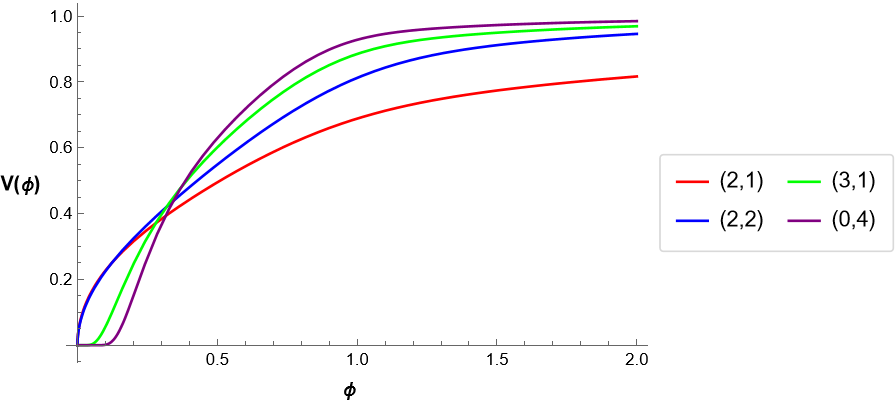}
\captionof{figure}{Graphic representation of the scalar potential for  $(m,n)$-models for different  values of the couple  $(m,n)$.  { We  have used units where $\kappa=\alpha=\beta=c=1$. }}
\label{F1}
\end{figure}
\end{center}

Inserting our  proposed function models  into the  above differential equation,  we  first  obtain  the following scalar coupling 
\begin{eqnarray}
\xi(\phi) = \lambda\int^{\kappa\phi}_0 e^{-\frac{\kappa^{m-1}}{\beta}(\frac{m\kappa^{n}}{m+n}x^{m+n}+x^m)}dx.
\end{eqnarray}
After integration  calculations, we find 
\begin{eqnarray}
V_1(\phi) &=&   e^{-\beta\frac{(\kappa\phi)^{2-(m+n)}\kappa}{\alpha m(m+n-2)} \; _2F_1(1,\frac{m+n-2}{n};\frac{m+2n-2}{n};-(\kappa\phi)^{-n})}\\
V_2(\phi) &= & c+\lambda \frac{8\kappa^4}{3\alpha^2}\int_{\kappa\phi}^{+ \infty}e^{-\big\lbrace\frac{\kappa^{m-1}}{\beta}(\frac{m\kappa^{n}}{m+n}x^{m+n}+x^m)+\beta\frac{(\kappa x)^{2-(m+n)}\kappa}{m(m+n-2)} \;_2F_1(1,\frac{m+n-2}{n};\frac{m+2n-2}{n};-(\kappa x)^{-n})\big\rbrace}dx.\notag\\
\end{eqnarray}
This provides the following   hypergeometric scalar potential 
\begin{eqnarray}
\small{V(\phi) = \frac{e^{-\beta\frac{(\kappa\phi)^{2-(m+n)}\kappa}{\alpha m(m+n-2)} \; _2F_1(1,\frac{m+n-2}{n};\frac{m+2n-2}{n};-(\kappa\phi)^{-n})}}{c+\lambda\frac{8\kappa^4}{3\alpha^2}\int_{\kappa\phi}^{+ \infty}e^{-\big\lbrace\frac{\kappa^{m-1}}{\beta}(\frac{m\kappa^{n}}{m+n}x^{m+n}+x^m)+\beta\frac{(\kappa x)^{2-(m+n)}\kappa}{m(m+n-2)} \;_2F_1(1,\frac{m+n-2}{n};\frac{m+2n-2}{n};-(\kappa x)^{-n})\big\rbrace}dx}}
\end{eqnarray}
where $_2F_1$  is the usual  hypergeometric function. It  is worth noting that  extra  constrains could be derived from the hypergeometric function behaviors and the physical requirements. Before going ahead,   
we illustrate the scalar potential   $V(\phi)$ for different values of $m$ and $n$ in Fig.(\ref{F1}) {  by using certain units   of  the involved parameters}.
The choice of $m$ and $n$ for this graphic  representation will be justified  in the coming discussions. It follows from this figure that the scalar potential becomes constant for large field values. Using Eqs (\ref{Eq3.1}) and (\ref{Eq3.2}), the slow-roll indices are  found to be 
\begin{eqnarray}
\epsilon_1 &=& \frac{\kappa^2\beta^2}{2\alpha m^2}\frac{1}{(\phi^n+1)^2\phi^{2m-2}},\\
\epsilon_2 &=& \frac{2\alpha m(m-1)\beta\phi^{-2m}\left(\phi^m((m-1)(\phi^n+1)+n\phi^n\right)-\beta\kappa^2\phi^2]}{2\alpha(1+\phi^n)^2m^2}.
\end{eqnarray}
Solving  the constraint $\epsilon_1(\phi_f)=1$, we obtain two quasi-homogeneous polynomials of the roots $\phi_f$ and $\phi_i$. Precisely,  the final
value of the scalar field    can be obtained via the following algebraic equation 
\begin{eqnarray}
(\phi_f^n+1)\phi^{m-1}_f-\frac{\kappa\beta}{m\sqrt{2\alpha}}=0.
\label{Eq4.9}
\end{eqnarray}
Using the  e-folding number  given by 
\begin{eqnarray}
N=-\int^{\phi_f}_{\phi_i}  \frac{A_1(\phi)}{A_2(\phi)} d\phi, 
\end{eqnarray}
we  find the   quasi-homogeneouss polynomial
\begin{eqnarray}
\phi_i^m(\frac{m}{m+n}\phi^n_i+1)-\frac{\kappa\beta}{m\sqrt{2\alpha}}\frac{(\frac{m}{m+n}\phi^n_f+1)}{\phi_f^n+1}\phi_f-N\beta=0.
\label{Eq4.11}
\end{eqnarray}
To avoid a discussion related to  the solution problems of polynomials,  we  could  impose constraints  on the couple  $(m,n)$ in order to furnish  consistent  inflationary models.   Combining the   quintic polynomial investigations with hypergeometric type behaviors,   we could consider the condition $ 2< n+m<5$   providing restrictions on the    $(m,n)$-models generating corroborated  findings having similar behaviors  of certain  results reported in  \cite{3}.  Additionally, one can clearly see that the model $(2,0)$ which has the error function as its scalar coupling function is omitted,  since the integral $\int^\infty_{\kappa\phi}  \frac{A_2(x)}{A_1(x)}  dx=\int^\infty_{\kappa\phi}\frac{1}{x}dx$ diverges.   To illustrate  the algorithm,   presented  here, we  treat the situations  corresponding to  $n\neq 0$. In what follows, the  discussion will be elaborated in terms of  the  relevant parameters defining  a moduli  space   $\cal M$    coordinated by  $\lbrace c,\alpha,\beta,N\rbrace$ by  fixing   $\kappa=1$.

\subsection{ Selected models}
Here, we  supply  a detailed  discussion on  some sets of  selected   $(m,n)$-models. Precisely,  we give the obtained  numerical values of the involved quantities including the ones associated with the  swampland criteria  program.
\subsubsection{$(2,1)$-models}
\begin{figure}[!ht]
\begin{center}
\centering
\begin{tabbing}
\centering
\hspace{8.cm}\=\kill
\includegraphics[width=8cm, height=7cm]{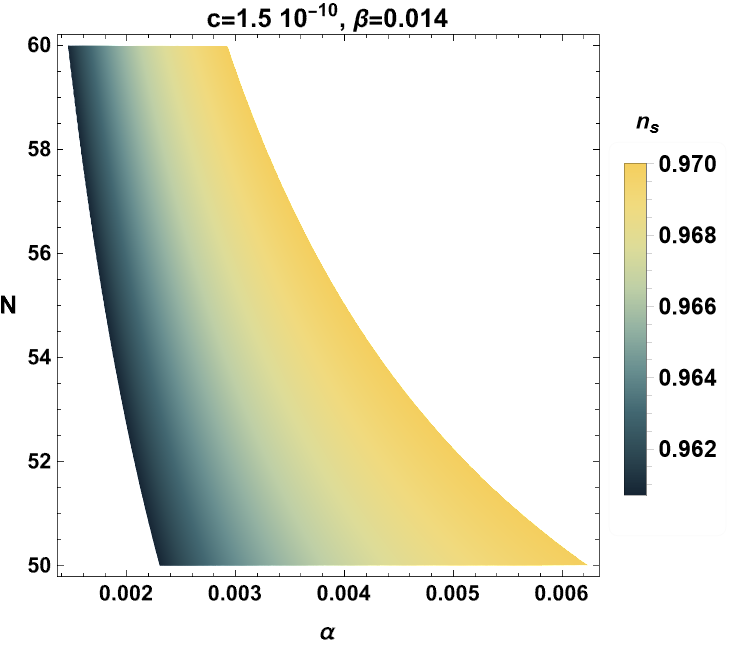}

\hspace{0.1cm} \includegraphics[width=8cm, height=7cm]{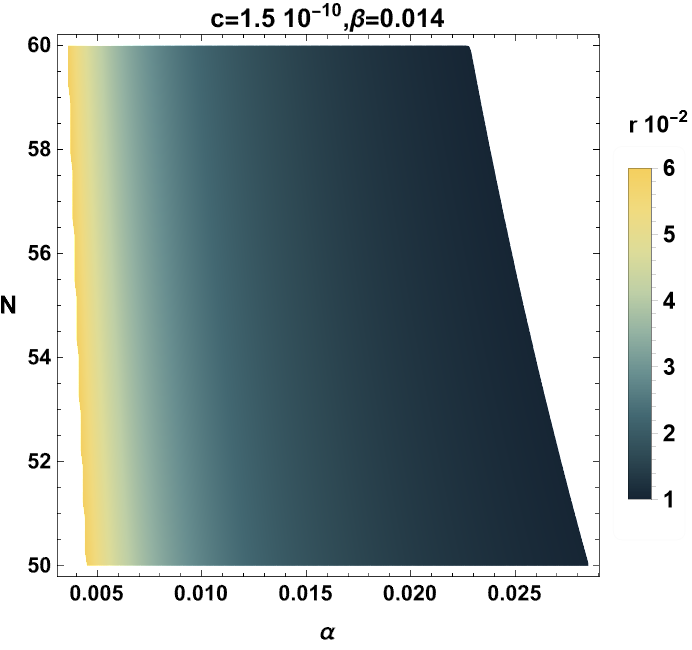}

  \end{tabbing}
\caption{Scalar spectral index $n_\tau$ (to the right) and the tensor to scalar ratio $r$ (to the left) in  terms of the e-folding number $N$ and the scaling parameter $\alpha$.}
\label{F2}
\end{center}
\end{figure}

\begin{figure}[!ht]
\begin{center}
\centering
\includegraphics[width=8cm, height=7cm]{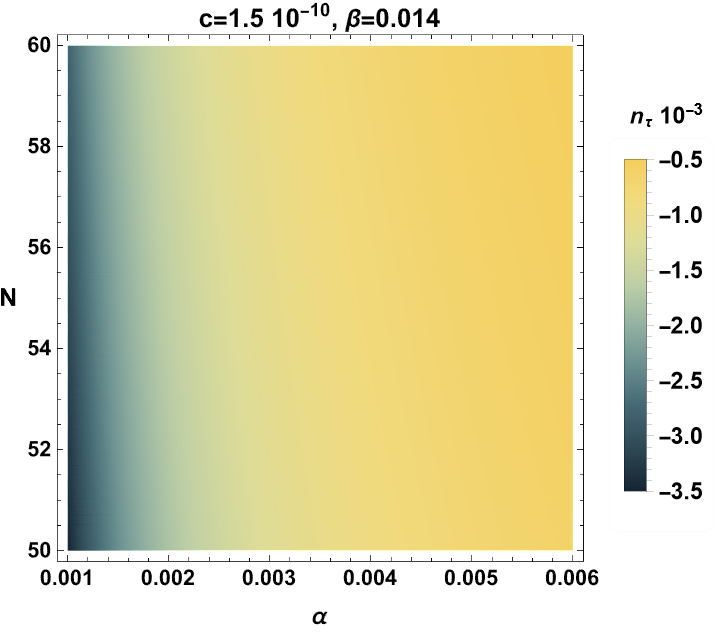}
\caption{Tensor spectral index $n_\tau$   by varying   the  e-folding number $N$ and the scaling parameter $\alpha$.}
\label{F3}
\end{center}
\end{figure}

\begin{figure}[!ht]
\begin{center}
\centering

\includegraphics[width=8cm, height=7cm]{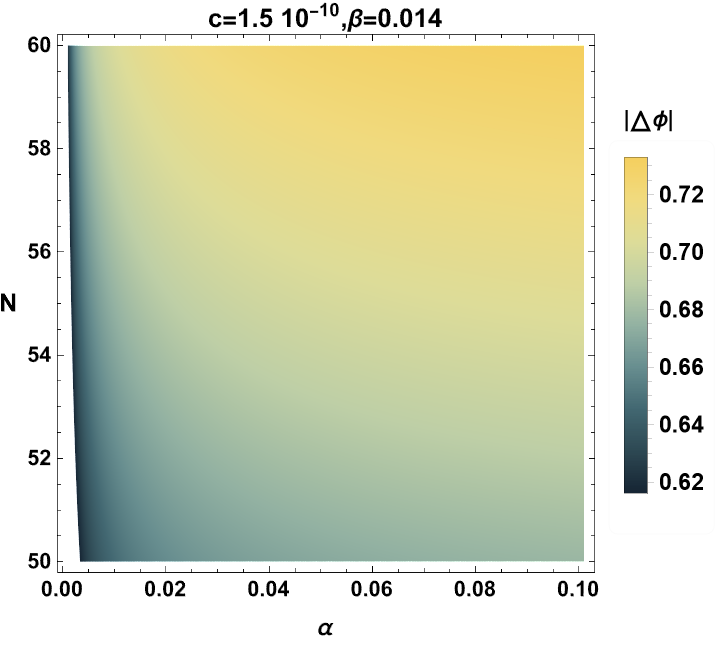}

\caption{The distance conjecture $\vert\Delta\phi\vert$ values   in terms  of the folding number $N$ and the scaling parameter $\alpha$.}
\label{F4}
\end{center}
\end{figure}

\begin{figure}[!ht]
\begin{center}
\centering
\begin{tabbing}
\centering
\hspace{8.cm}\=\kill
\includegraphics[width=8cm, height=7cm]{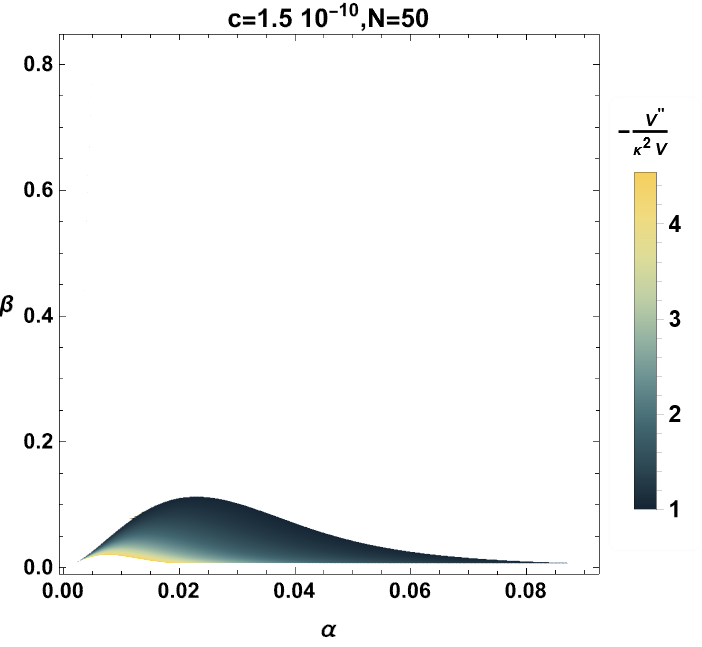}

\hspace{0.1cm} \includegraphics[width=8cm, height=7cm]{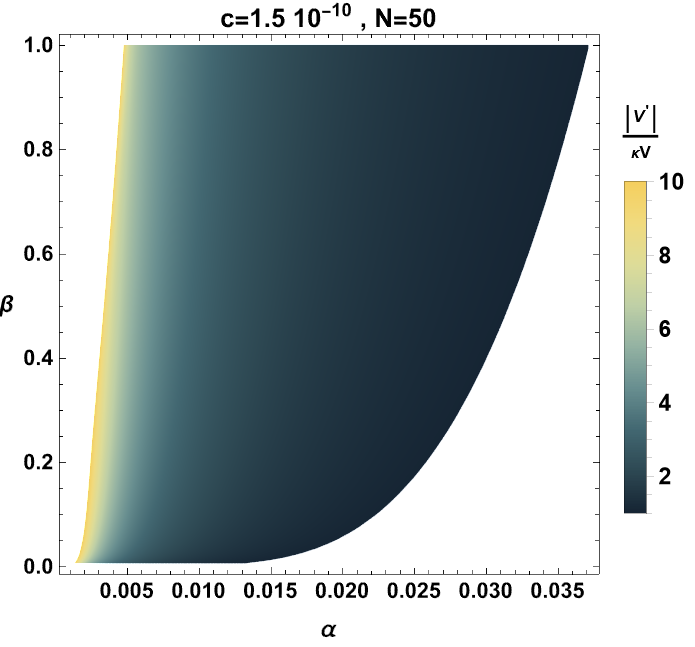}

  \end{tabbing}
\caption{Tachyonic instability conjecture  $-\frac{V^{\prime\prime}}{\kappa^2 V}$ (at the  left)  and the runaway instability conjecture  $\frac{\vert V^\prime\vert}{\kappa V}$ (at the right)    as  functions of the free parameter $\beta$ and the scaling parameter $\alpha$.}
\label{F5}
\end{center}
\end{figure}

To obtain only positive values of the  scalar field at the end of the inflationary era, the solution of Eq.(\ref{Eq4.9})  has the following form 
\begin{equation}
  \phi_f=\frac{1}{2}(\sqrt{1+\frac{\kappa\beta}{\sqrt{2\alpha}}}-1),
   \end{equation}  
   where one has used $\beta>0$.
 To remove the imaginary initial scalar  field values,  we consider only the real solution of  Eq.(\ref{Eq4.11}) being
 \begin{equation}
 \phi_i=\frac{\psi^2-\psi+1}{2\psi},
 \end{equation}
 were one has 
 \begin{eqnarray}
  \psi &=& \Big[6(\frac{\kappa\beta}{2\sqrt{2\alpha}}\frac{(\frac{2}{3}\phi_f+1)}{\phi_f+1}\phi_f+N\beta)\notag\\&+&2\sqrt{-3\Big(\frac{\kappa\beta}{2\sqrt{2\alpha}}\frac{(\frac{2}{3}\phi_f+1)}{\phi_f+1}\phi_f+N\beta)\Big)+9\Big(\frac{\kappa\beta}{2\sqrt{2\alpha}}\frac{(\frac{2}{3}\phi_f+1)}{\phi_f+1}\phi_f+N\beta)\Big)^2}-1\Big]^\frac{1}{3}.\notag \\
 \end{eqnarray}
 For this model, we should select a convenient point in the moduli space $\cal M$.  The  values of the relevant  cosmological quantities $n_S$, $r$ and $ n_\tau$   will be investigated by varying the e-folding number $N$ and the rescaling parameter $\alpha$. 
Concretely, we consider the point $P_1\equiv(c=1.5\ 10^{-10},N=50,\beta=0.014,\alpha=0.0045)$. 
 The corresponding value for the scalar spectral index is $n_S=0.967956543$, and the scalar to tensor ratio is $r=0.0629716367$. For generic points of $\mathcal{M}$, acceptable values  of the scalar index and the scalar to tensor ratio are illustrated in the left and the right of Fig.(\ref{F2}), respectively. In certain regions of the  moduli space,  the obtained  results match 
perfectly with the Planck constraints, where the corresponding values are represented by an ivory tan tiled color. For the point $P_1$,  moreover,  we find that the value of the tensor index is $n_\tau=-0.00787145644$. The stringy constraint $r\approx -\Delta n_\tau$ is also verified, where we   obtain $\frac{r}{n_\tau}=-7.99999809$ and $\Delta=-(8+64\frac{H}{\alpha n_\tau}\xi^\prime)=8$.  Certain cosmological  values corresponding to generic points of  the moduli space $\mathcal{M}$ are represented in Fig.(\ref{F3}). In such a figure, the  values of the tensor spectral index $n_\tau$    are provided by taking the different values of the  e-folding number $N$ and the scaling parameter $\alpha$.\\
Concerning the swampland criteria,  the relevant quantities $\frac{\vert V^\prime\vert}{\kappa V}$ and $-\frac{V^{\prime\prime}}{\kappa^2 V}$ are  discussed  by varying of the free parameter $\beta$ and the rescaling parameter $\alpha$. Indeed,   the initial and the final values of the field which are found to be $\phi_i=0.694217026$ and $\phi_f=0.0690223947$ providing $\Delta\phi=-0.625194609$ being in accordance with the distance conjecture. This value lays in the dark teal region of Fig.(\ref{F4}). For the de Sitter conjectures, we get values consistent with the swampland criteria. For the runway instability, we obtain $\frac{\vert V^\prime\vert}{\kappa V}=3.35233879$, located at the dark teal region illustrated in the right of Fig.(\ref{F5}). The tachyonic instability is found to be $-\frac{V^{\prime\prime}}{\kappa^2 V}=5.56149721$, laying at the ivory tan region being represented in the left of Fig.(\ref{F5}).  In the present  rescaled gravity model,  we have obtained  values  greater than $O(1)$ showing that   one has  acceptable  conditions.\\
The chosen range for the free parameter $\beta$ is motivated by the fact that the rescaling parameter $\alpha$ is at most equal to one, according to the previous discussion. Following to a numerical analysis for $\beta>>\mathcal{O}(1)$, we get $\Delta\phi\sim\phi_f>>\mathcal{O}(1)$. To avoid large field values, we consider the following choice $0<\beta\leq\mathcal{O}(1)$. One should note that the main objective is not just to meet observational data, but also to extract the swampland regions in the moduli space. The arbitrary values of such free parameters could produce  models which could be checked for consistency with quantum gravity being a basic approach of the  present investigation.

We also highlight the numerical values of the inflationary indices, which are as follows $\epsilon_1=0.00393572729$, $\epsilon_2=0.00815027580$, $\epsilon_3=5.74050693\ 10^{-16}$ and $\epsilon_4=8.71408223\ 10^{-10}$. The sound velocity is practically $c_A=1.000000$ which is consistent with the absence of instability. The scalar potential  value for the obtained set of the slow-roll indices is $V(\phi)=1.66407053\ 10^{-9}$, where the order of the potential seems to be related to the integration constant via the relation  $\log_{10}(V(\phi))\simeq \log_{10}(c)-1$.\\

Via a qualitative analysis,  we can  observe that the landscape is surrounded  by a vast swampland of consistent-looking semi-classical EFTs corresponding to  the union of all the moduli space regions verifying the conditions checked above. The presence of  such  a  vast  behavior  indicates  a degree of consistency of the EFTs with quantum gravity.

\subsubsection{$(1,2)$-model}
\begin{figure}[!ht]
\begin{center}
\centering
\begin{tabbing}
\centering
\hspace{8.cm}\=\kill
\includegraphics[width=8cm, height=7cm]{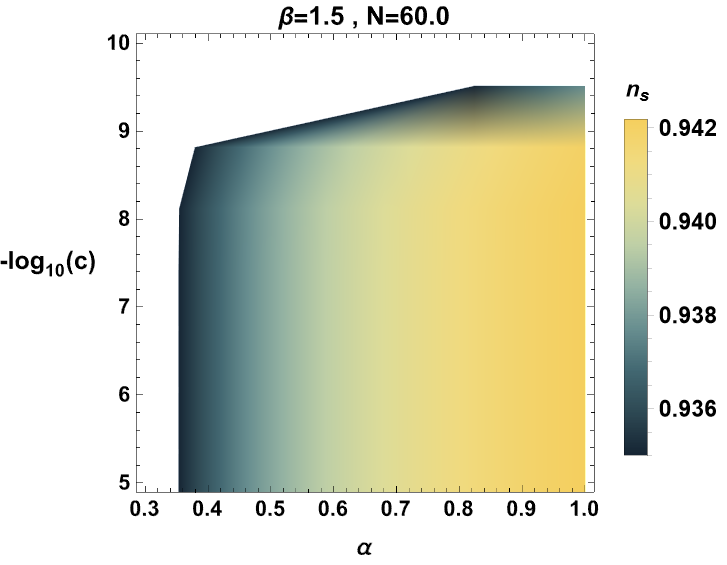}
\hspace{0.1cm} \includegraphics[width=8cm, height=7cm]{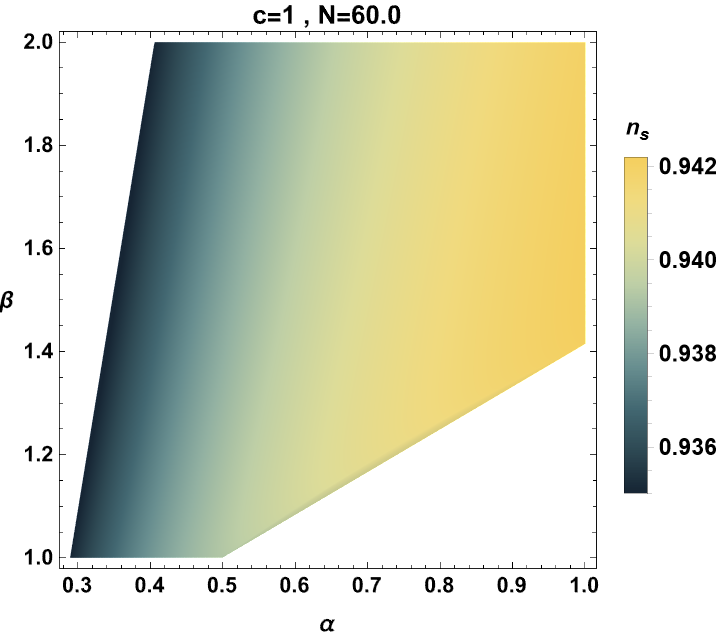}
  \end{tabbing}
\caption{ Right: Scalar scalar index $n_s$ in  terms  of the free parameter $\beta$ and scaling parameter $\alpha$. Left:  Scalar scalar index $n_s$ in  terms of the  integration constant $c$ and the scaling parameter $\alpha$.}
\label{F6}
\end{center}
\end{figure}

In this part, we consider the   $(1,2)$-model.    In particular, we  use   similar computations    performed  in the previous model.  Indeed, the solution   of Eq.(\ref{Eq4.9}),  giving  positive field values,  is  
\begin{equation}
  \phi_f=\sqrt{\frac{\kappa\beta}{\sqrt{2\alpha}}-1}.
   \end{equation}  
 For this model, the  real  solution of  Eq.(\ref{Eq4.11})  is
 \begin{equation}
 \phi_i=\frac{\psi^2-2^\frac{2}{3}}{2^\frac{1}{3}\psi}
 \end{equation}
 where one has used
 \begin{equation}
\psi=3\Big(\frac{\kappa\beta}{\sqrt{2\alpha}}\frac{(\frac{1}{3}\phi^2_f+1)}{\phi_f^2+1}\phi_f+N\beta)+\sqrt{4+9(\frac{\kappa\beta}{\sqrt{2\alpha}}\frac{(\frac{1}{3}\phi^2_f+1)}{\phi_f^2+1}\phi_f+N\beta))^2}\Big)^\frac{1}{3}. 
 \end{equation}
 To get the numerical values of the relevant quantities,  one should consider a point in the moduli space     $\cal M$  with the reduced units.  In particular, we take  the point   $P_2\equiv (c=1,N=60,\beta=1.5,\alpha=0.9)$ of     $\cal M$.   Indeed, the values for the scalar 
spectral index  and the tensor-to-scalar ratio are  $n_S=0.941780150$ and  $r=0.0380603224$,  respectively.
This model is  incompatible with the Planck data where the maximum value of the scalar spectral index seems to be around the point $P_2$,  as detected in Fig.(\ref{F6}). 

 In the left panel of Fig.(\ref{F6}), the  values of $n_S$ are represented in the colored bar by varying the integration constant $c$ and the rescaling parameter $\alpha$. In the right panel of Fig.(\ref{F6}), the  values of $n_S$ are represented by varying the free parameter $\beta$ and the rescaling parameter $\alpha$. It  follows from this figure that the   maximal value of $n_S$  lays in the regions defined by the constraints  $0.9\leq\alpha<1$, $1.4<\beta<1.8$ and $-Log_{10}(c)<8.5$.  However, the scalar to  the tensor ratio and the tensor spectral index are found to be  $r=0.0380603224$ and  $n_\tau=-0.0047575403$, respectively.  These two last values match perfectly with  the observational data. \\
 
 { The illustrated interval of $-Log_{10}(c)$ is constrained by the mentioned conditions on $\beta$ and $\alpha$ parameters. The regions that are not shown are not relevant. For $-Log_{10}(c)>10$,  the values of $n_S$ are not even close to acceptable ones. For $-Log_{10}(c)<8$, moreover, the values of $n_S$ are independent of $c$.}
 
 The incompatibility of the scalar spectral index $n_S$ with the observational data  allows   one  to rule out such a model without moving into   the  swampland criteria  analysis.
\subsubsection{$(2,2)$-model}
\begin{figure}[!ht]
\begin{center}
\centering
\includegraphics[width=8cm, height=7cm]{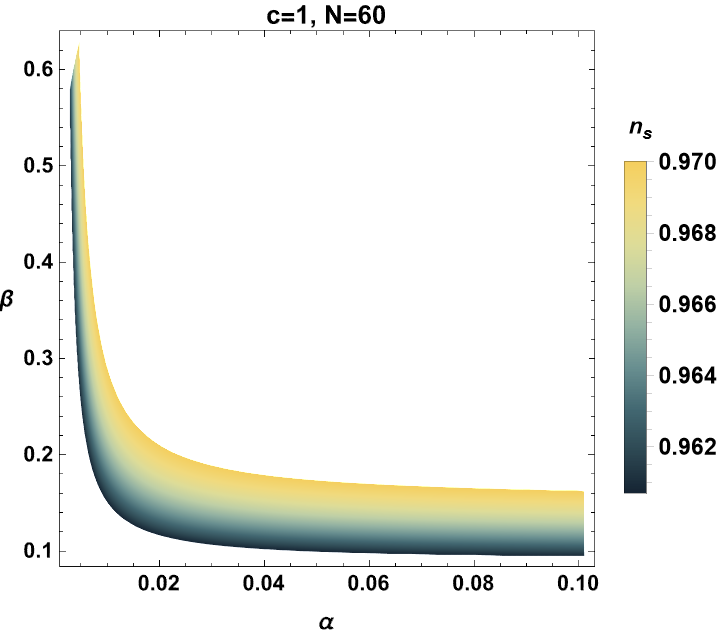}
\caption{Scalar spectral index $n_S$ values  as a  function of the free parameter $\beta$ and the scaling parameter $\alpha$.}
\label{F7}
\end{center}
\end{figure}

For this model, we find   that the final value of the scalar field  is 
\begin{equation}
  \phi_f=\frac{\psi^2-6^\frac{1}{3}}{18^\frac{1}{3}\psi},
   \end{equation}  
 where we have used 
\begin{equation}
   \psi=\Big(9\frac{\kappa\beta}{\sqrt{2\alpha}}+\sqrt{12+81(\frac{\kappa\beta}{\sqrt{2\alpha}})^2}\Big)^\frac{1}{3}.
   \end{equation}   
The real and positive solution of Eq.(\ref{Eq4.11}) is
 \begin{equation}
 \phi_i=\sqrt{\sqrt{1+2\Big(\frac{\kappa\beta}{2\sqrt{2\alpha}}\frac{(\frac{1}{2}\phi^2_f+1)}{\phi_f^2+1}\phi_f+N\beta)\Big)}-1}.
 \end{equation}
Now, we  consider the  point   $P_3\equiv(c=1,N=60,\beta=0.13,\alpha=0.015)$ of  the moduli space $\mathcal{M}$.   At such a  point,  we find  that the scalar spectral index is $n_S =0.961151063$.  The scalar to tensor ratio is found to be  $r=0.0592807233$. For generic points of $\mathcal{M}$,  the  obtained values  of the scalar index are illustrated in Fig.(\ref{F7}). 

 A close examination shows that one has found acceptables values   matching  
perfectly with the Planck constraints, where the values  corresponding to $n_S$  are  represented by a dark teal color.

\begin{figure}[!ht]
\begin{center}
\centering
\includegraphics[width=8cm, height=7cm]{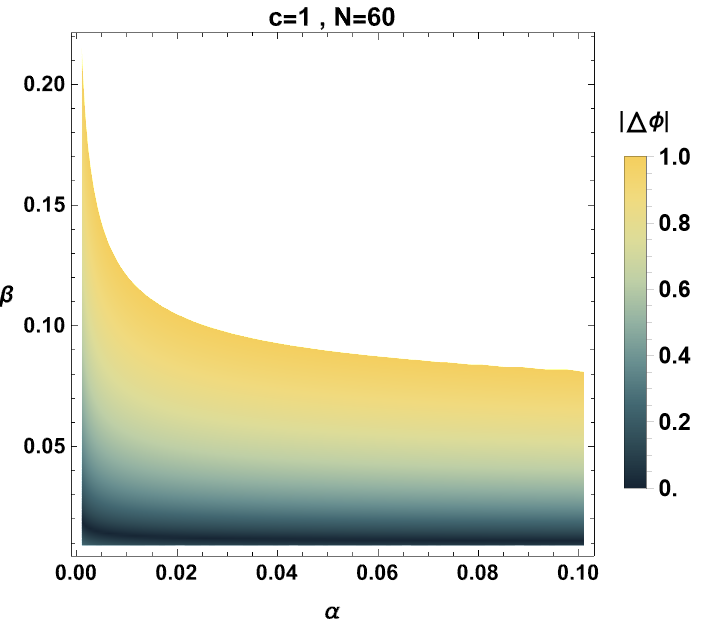}

\caption{Distance conjecture $\vert\Delta\phi\vert$ values  as a function of the free parameter $\beta$ and the scaling parameter $\alpha$.}
\label{F8}
\end{center}
\end{figure}

After computations, we find that the numerical value of the tensor index is $n_\tau=-0.0074100904$.   It has been remarked that the  stringy constraint $r\approx -\Delta n_\tau$ is also  satisfied, where we   obtain   $\frac{r}{n_\tau}=-8.00000000$ leading to   $\Delta=8$.
 \\
Regarding the swampland criteria, we  investigate the associated conditions. Indeed, we  can first  calculate    the initial and the final values of the  scalar field. They are found to be    $\phi_i=1.65252721$ and $\phi_f=0.568406165$, respectively.   Such  values  provide   $\Delta\phi=1.08412099$ showing an  inconsistent  behavior with the distance conjecture.  The constraints on the scalar index $n_S$ and the distance conjecture $\vert\Delta\phi\vert<1$  are satisfied  separately. The values that are bounded by the distance conjecture of $\vert\Delta\phi\vert$ are represented in Fig.(\ref{F8}).

In this figure,  the   obtained  values of the relevant quantity  $\vert\Delta\phi\vert$ are represented  in the colored bar by varying the free parameter $\beta$ and the rescaling parameter $\alpha$. The discrepancy between the two conditions is apparent in the two previous  figures, where it is clear that the two consistency regions never intersect. For the de Sitter conjectures, we get values consistent with the swampland criteria. For the runway instability, we obtain $\frac{\vert V^\prime\vert}{\kappa V}=1.04889643$ while the tachyonic instability is found to be $-\frac{V^{\prime\prime}}{\kappa^2 V}=1.86980057$. In the present  rescaled gravity model,  we obtain   values  greater than $O(1)$  revealing that one has  acceptable  conditions.

\subsection{Discussion on the remaining $(m,n)$-models}
Instead of repeating the  above treatment, we prefer to provide the relevant results  of the remaining models.  We start by noting that   the  $(3,1)$-model  should be   excluded, due to the absence of  the  real initial field values.  To see that,  we consider the solution
\begin{equation}
\phi_i=-\frac{1}{3}+\frac{\psi}{3\sqrt{2}}+\frac{1}{2}\sqrt{\frac{2\psi^3+4\psi-8\sqrt{2}}{9\sqrt{\psi}}}.
\end{equation}
In this equation,  we have used 
\begin{eqnarray}
\psi &=&( 2-\frac{ 2^\frac{1}{3}6\rho}{\Pi}+3\Pi)^{\frac{1}{2}}\\
\rho &=& \frac{\beta}{\sqrt{2\alpha}}\frac{\frac{3}{4}\phi_f+1}{\phi_f+1}\phi_f+N\beta
\end{eqnarray}
where  one  has  $\Pi=\Pi(\beta,\alpha,N)=\big(\sqrt{\rho^2+4\rho^3}-\rho\big)^\frac{1}{3}$. For the solution to be real,   one  must have $\psi^2>0$. However,  it  has been shown that this is contradictory since one  has the condition $\sqrt{\rho^2+4\rho^3}-\rho>0$.  Similar behaviors have been remarked for the  $(1,3)$-model.  In fact,    Eq.(\ref{Eq4.11})  has only two considerable solutions. Explicitly, we  have  
\begin{eqnarray}
\label{sol}
\phi^{++}_i=\frac{\psi}{\sqrt{2}}+\frac{1}{2}\sqrt{2\psi^2-\frac{4\sqrt{2}}{\psi}},  \nonumber\\
\phi^{+-}_i=\frac{\psi}{\sqrt{2}}-\frac{1}{2}\sqrt{2\psi^2-\frac{4\sqrt{2}}{\psi}}.
\end{eqnarray}
It is worth noting that  two  other solutions  generate negative field values,  which are omitted by  physical constraints.  We see clearly that the equation $\psi^3-2\sqrt{2}=0$ has the real solution $\sqrt[3]{2\sqrt{2}}$ with multiplicity equal to 3.  The first solution  of Eq.(\ref{sol}) gives relatively large field values with the condition $\beta>\sqrt{2\alpha}$.  This results in   vanishingly small values of  the scalar coupling derivative $\xi^\prime(\phi)$, being omitted due  to  analytical arguments. For the second solution,  we basically get completely inconsistent results with the Planck data. Thus,  this model should be  ruled out.

Lastly,   we consider  the  $(4,0)$-model.   It is worth noting that    the associated coupling constant is not considered as a free parameter.   Using the boundary conditions on the scalar coupling function,  we find that $\lambda=\lambda(\beta)=\xi^\prime(0)$ meaning that this quantity is not independent of the model  free parameters. In fact, it provides a reduced   moduli space. In  Tab. (\ref{Tab1}), we collect the obtained numerical values of the dealt with quantities and the swampland criteria.

\begin{table}[!h]
\begin{center}
$\begin{tabular}{|c|c|}
\hline
& $(4,0)$-model for the point $ P_4\equiv(c=1,N=50,\beta=\frac{2}{35},\alpha=0.004)$ \\
\hline
 $n_{S}$ &  $0.963496923$ \\
\hline
$r$ &  $0.0576666035$ \\
\hline
$n_{\tau}$ &  $-0.00720832543$ \\
\hline
$\vert \Delta\phi\vert$ &  $0.669151545$ \\
\hline
$\frac{\vert V\vert}{\kappa V}$ &  $1.34241629$ \\
\hline
$-\frac{V^{\prime\prime}}{\kappa2V}$ &  $1.85976577$ \\
\hline
\end{tabular}$
\caption{ Numerical values of the  $(4,0)$-model for the point $ P_4\equiv(c=1,N=50,\beta=\frac{2}{35},\alpha=0.004)$.}
\label{Tab1}
\end{center}
\end{table}

For the chosen  moduli space point,   it  has been  observed for this table that the obtained   numerical values are acceptable for  phenomenological interest. They    could be corroborated by certain observational data. 

\section{Conclusions and discussions}

  Inspired by differential equations  involving  special functions, we  have presented  an algorithm   allowing one to get  new   hypergeometric type scalar potentials from  the stringy  correction   function  $\xi=\xi(\phi)$, coupled to the Gauss-Bonnet term.   In particular, we have constructed  and analyzed  certain models of phenomenological interest by providing  inflationary  predictions of the relevant cosmological observables.    In particular, we have furnished   a  family models, refereed  to as $(m,n)$-models where the  couple   $(m,n)$  is constrained from    the  hypergeometric   potential behaviors and certain  physical arguments. We have shown  that these  models  provide corroborated findings.  Using the falsification scenario,  we have   confronted  the  predictions  brought by the obtained  models with the Planck  observational data for  such a   stringy rescaled gravity.  Then, we   have  approached   the corresponding  swampland conjectures.  Among others, we  have found  that   the swampland criteria are satisfied for small values of the  slow-roll   parameters.

The present work comes up with  some sets of  questions.   A natural question  concerns  the implementation of extra stringy corrections including the  kinetic Einstein coupling.   It is possible that such  contributions  could be useful to produce consistent models via the swampland program. The second question may concern the naturalness  arguments of the proposed  physical theories.  Such questions could be addressed in future  investigations.

\textbf{Data availability statements}:   Data sharing is not applicable to this article.

\textbf{Acknowledgements}:   Saad Eddine Baddis   would like to thank  H. Belmahi  for comments on an early draft of this work.   He thanks   the organizers of the String-Math 2024, the Abdus Salam International Centre for Theoretical Physics (ICTP),
Trieste, Italy, for hospitality. He would like also  to thank C. Vafa for discussions and   for  certain clarifications at ICTP.

\end{document}